\documentstyle[epsfig,12pt]{article}

\def\lord{$ \raisebox{-.3ex}{$\stackrel{<}{_{\sim}}$} $}

\title{\begin{flushright}
{\normalsize NUC-MINN-2001/11-T\\
September 2001 \\}
\end{flushright}
\vspace*{0.3in}
{\bf HIGH TEMPERATURE MATTER AND GAMMA RAY SPECTRA FROM MICROSCOPIC BLACK 
HOLES}}
\author{{R. G. Daghigh and J. I. Kapusta} \vspace*{0.1in}\\
 {\it School of Physics and Astronomy, University of Minnesota}\\
 {\it Minneapolis, MN 55455, USA}}

\date{}

\begin{document}

\maketitle

\begin{abstract}

The relativistic viscous fluid equations describing the outflow of high 
temperature matter created via Hawking radiation from microscopic black holes 
are solved numerically for a realistic equation of state.  We focus on black 
holes with initial temperatures greater than 100 GeV and lifetimes less than 6 
days.  The spectra of direct photons and photons from $\pi^0$ decay are 
calculated for energies greater than 1 GeV.  We calculate the diffuse gamma ray 
spectrum from black holes distributed in our galactic halo.  However, the most 
promising route for their observation is to search for point sources emitting 
gamma rays of ever-increasing energy. 

\end{abstract}

\section{Introduction}

Hawking radiation from black holes \cite{Hawk} is of fundamental interest
because it relies on the application of relativistic quantum field theory in the
presence of the strong field limit of gravity, a situation that could 
potentially be observed.  It is also of great interest because of the 
temperatures involved.  A black hole with mass $M$ radiates thermally with a 
Hawking temperature $T_H = m_{\rm P}^2/8\pi M$
where $m_{\rm P} = G^{-1/2} = 1.22\times 10^{19}$ GeV is the Planck mass.
(Units are $\hbar = c = k_{\rm B} = 1$.)  In order for the black hole
to evaporate rather than accrete it must have a temperature greater than that of 
the present-day black-body radiation of the universe of 2.7 K = 2.3$\times 10^{-
4}$ eV.  This implies that $M$ must be less than $1\%$ of the mass of the Earth.
Such small black holes most likely would have been formed primordially; there is 
no other mechanism known to form them.  As the black hole radiates, its mass 
decreases and its temperature increases until $T_H$ becomes comparable to the
Planck mass, at which point the semi-classical calculation breaks down and the 
regime of full quantum gravity is entered.  Only in two other situations are 
such enormous temperatures achievable: in the early universe and in central 
collisions of heavy nuclei like gold or lead.  Even then only about $T = 500$ 
MeV is reached at the RHIC (Relativistic Heavy Ion Collider) just completed at
Brookhaven National Laboratory and $T = 1$ GeV is expected at the LHC (Large
Hadron Collider) at CERN to be completed in 2006.  Supernovae and newly formed
neutron stars only reach temperatures of a few tens of MeV.
To set the scale from fundamental physics, we note that the spontaneously broken
chiral symmetry of QCD gets restored in a phase transition/rapid crossover at a
temperature around 170 MeV, while the spontaneously broken gauge symmetry in the
electroweak sector of the standard model gets restored in a phase
transition/rapid crossover at a temperature around 100 GeV.  The fact that
temperatures of the latter order of magnitude will never be achieved in a
terrestrial experiment should motivate us to study the fate of microscopic black
holes during the final days, hours and minutes of their lives when their 
temperatures have risen to 100 GeV and above.  In this paper we shall focus on 
Hawking temperatures greater than 100 GeV.  The fact that microscopic black 
holes have not yet been observed \cite{review} should not be viewed as a 
deterrent, but rather as a challenge for the new millennium!

There is some uncertainty over whether the particles scatter from each
other after being emitted, perhaps even enough to allow a fluid description of
the wind coming from the black hole.  Let us examine what might happen as the
black hole mass decreases and the associated Hawking temperature increases.

When $T_H \ll m_{\rm e}$ (electron mass) only photons, gravitons,
and neutrinos will be created with any significant probability.  These
particles will not interact with each other but will be emitted into
the surrounding space with the speed of light.
Even when $T_H \approx m_{\rm e}$ the Thomson cross section is too
small to allow the photons to scatter very frequently in the rarified
electron-positron plasma around the black hole.  This may change when
$T_H \approx 100$ MeV when muons and charged pions are created in
abundance.  At somewhat higher temperatures hadrons are copiously produced and
local thermal equilibrium may be achieved, although exactly how is an unsettled
issue.  Are hadrons emitted directly by the black hole?  If so, they will be
quite abundant at temperatures of order 150 MeV because their mass spectrum
rises exponentially (Hagedorn growth as seen in the Particle Data Tables
\cite{PDG}).  Because they are so massive they move nonrelativistically and may
form a very dense equilibrated gas around the black hole.  But hadrons are
composites of quarks and gluons, so perhaps quarks and gluon jets are emitted
instead?  These jets must decay into the observable hadrons on a typical
proper length scale of 1 fm and a typical proper time scale of 1 fm/c.
This was first studied by
MacGibbon and Webber \cite{MW} and MacGibbon and Carr \cite{MC}.  Subsequently
Heckler \cite{Ha} argued that since the emitted quarks and gluons are so densely
packed outside the event horizon they are not actually fragmenting into
hadrons in vacuum but in something more like a quark-gluon plasma, so perhaps
they thermalize.  He also argued that QED bremsstrahlung and pair production
were sufficient to lead to a thermalized QED plasma when $T_H$ exceeded 45 GeV
\cite{Hb}.  These results are somewhat controversial and need to be confirmed.
The issue really is how to describe the emission of wavepackets via the Hawking
mechanism when the emitted particles are (potentially) close enough to be
mutually interacting.  A more quantitative treatment of the particle
interactions on a semiclassical level was carried out by Cline, Mostoslavsky and
Servant \cite{cline}.  They solved the relativistic Boltzmann equation with QCD
and QED interactions in the relaxation-time approximation. It was found that
significant particle scattering would lead to a photosphere though not perfect
fluid flow.

Rather than pursuing the Boltzmann transport equation one of us applied 
relativistic viscous fluid equations to the problem assuming sufficient particle 
interaction \cite{me}. It was found that a self-consistent description emerges 
of a fluid just marginally kept in local thermal equilibrium, and that viscosity 
is a crucial element of the dynamics.  The purpose of this paper is a more 
extensive analysis of these equations and their observational consequences.  The 
plan is as follows.  In section 2 we give a brief review of Hawking radiation 
sufficient for our uses.  In section 3 we give the set of relativistic viscous 
fluid equations necessary for this problem along with the assumptions that go 
into them.  In section 4 we suggest a relatively simple parametrization of the 
equation of state for temperatures ranging from several MeV to well over 100 
GeV.  We also suggest a corresponding parametrization of the bulk and shear 
viscosites.  In section 5 we solve the equations numerically, study the scaling 
behavior of the solutions, and check their physical self-consistency.
In section 6 we estimate where the transition from viscous fluid flow to
free-streaming takes place.  In section 7 we calculate the instantaneous and time-integrated spectra of high energy photons from the two dominant sources: direct and neutral pion decay.  In section 8 we study the diffuse gamma ray spectrum from microscopic black holes distributed in our galactic halo.  We also study the systematics of gamma rays from an individual black hole, should we be so fortunate to observe one.  We conclude the paper in section 9.

\section{Hawking Radiation}

There are at least two intuitive ways to think about Hawking radiation from
black holes.  One way is vacuum polarization.  Particle-antiparticle pairs are
continually popping in and out of the vacuum, usually with no observable effect.
In the presence of matter, however, their effects can be observed.  This is the
origin of the Lamb effect first measured in atomic hydrogen in 1947.  When pairs
pop out of the vacuum near the event horizon of a black hole one of them may be
captured by the black hole and the other by necessity of conservation laws will
escape to infinity with positive energy.  The black hole therefore has lost
energy - it radiates.  Due to the general principles of thermodynamics applied
to black holes it is quite natural that it should radiate thermally.  An
intuitive argument that is more quantitative is based on the uncertainty
principle.  Suppose that we wish to confine a massless particle to the vicinity
of a black hole.  Given that the average momentum of a massless particle at
temperature $T$ is approximately $\pi T$, the uncertainty principle requires
that confinement to a region the size of the Schwarzschild diameter places a
restriction on the minimum value of the temperature.
\begin{equation}
\pi T \cdot 2 r_S > 1/2
\end{equation}
The minimum is actually attained for the Hawking temperature.  The various
physical quantities are related as $r_S = 2M/m_{\rm P}^2 =
1/4\pi T_H$.

The number of particles of spin $s$ emitted with energy $E$ per unit time is
given by the formula
\begin{equation}
\frac{dN_s}{dEdt} = \frac{\Gamma_s}{2\pi} \,
\frac{1}{\exp(E/T_H)-(-1)^{2s}} \, .
\end{equation}
All the computational effort really goes into calculating the absorption
coefficient $\Gamma_s$ from a relativistic wave equation in the presence of a
black hole.  Integrating over all particle species yields the luminosity.
\begin{equation}
L = -\frac{dM}{dt} = \alpha(M) \frac{m_{\rm P}^4}{M^2} =
64 \pi^2 \alpha(T_H) T_H^2 \, .
\end{equation}
Here $\alpha(M)$ is a function reflecting the species of particles available
for creation in the gravitational field of the black hole.  It is generally
sufficient to consider only those particles with mass less than $T_H$;
more massive particles are exponentially suppressed by the Boltzmann factor.
Then
\begin{equation}
\alpha = 2.011\times 10^{-8} \left[ 4200 N_0 + 2035 N_{1/2} + 835 N_1 + 95 N_2
\right]
\end{equation}
where $N_s$ is the net number of polarization degrees of freedom for all
particles with spin $s$ and with mass less than $T_H$.  The coefficients for
spin 1/2, 1 and 2 were computed by Page \cite{Page} and for spin 0 by Sanchez
\cite{Sanchez}.  In the standard model $N_0 = 4$ (Higgs), $N_{1/2} = 90$ (three
generations of quarks and leptons), $N_1 = 24$ (SU(3)$\times$SU(2)$\times$U(1)
gauge theory), and $N_2 = 2$ (gravitons).  This assumes $T_H$ is greater than
the temperature for the electroweak gauge symmetry restoration.
Numerically $\alpha(T_H > 100 \,{\rm GeV}) = 4.43\times 10^{-3}$.  Starting with
a black hole of temperature $T_H$, the time it takes to evaporate/explode is
\begin{equation}
\Delta t = \frac{m_{\rm P}^2}{3 \alpha(T_H) (8\pi T_H)^3} \, .
\end{equation}
This is also the characteristic time scale for the rate of change of the
luminosity of a black hole with temperature $T_H$.

At present a black hole will explode if $T_H > 2.7$ K and correspondingly
$M < 4.6\times 10^{25}$ g which is approximately 1\% of the mass of the Earth.
More massive black holes are cooler and therefore will absorb more matter and
radiation than they radiate, hence grow with time.  Taking into account emission
of gravitons, photons, and neutrinos a critical mass black hole today has a
Schwarszchild radius of 68 microns and a lifetime of $2\times10^{43}$ years.

\section{Relativistic Viscous Fluid Equations}

The relativistic
imperfect fluid equations describing a steady-state, spherically symmetric flow
with no net baryon number or electric charge and neglecting gravity
(see below) are $T^{\mu\nu}_{\;\;\;\;;\nu} =$ {\em black hole source}.  The
nonvanishing components of the energy-momentum tensor in radial coordinates are
\cite{MTW}
\begin{eqnarray}
T^{00}&=& \gamma^2 (P+\epsilon) -P + v^2 \Delta T_{\rm diss} \nonumber \\
T^{0r}&=& v\gamma^2 (P+\epsilon) + v \Delta T_{\rm diss} \nonumber \\
T^{rr}&=& v^2\gamma^2 (P+\epsilon) +P + \Delta T_{\rm diss}
\end{eqnarray}
representing energy density, radial energy flux, and radial momentum flux,
respectively, in the rest frame of the black hole.  Here
$v$ is the radial velocity with $\gamma$ the corresponding Lorentz factor, $u =
v\gamma$, $\epsilon$ and $P$ are the local energy density and pressure, and
\begin{equation}
\Delta T_{\rm diss} = -\frac{4}{3}\eta \gamma^2 \left( \frac{du}{dr}
-\frac{u}{r}\right) - \zeta \gamma^2 \left( \frac{du}{dr}
+\frac{2u}{r}\right) \, ,
\end{equation}
where $\eta$ is the shear viscosity and $\zeta$ is the bulk viscosity.  A
thermodynamic identity gives $Ts = P + \epsilon$ for zero chemical potentials,
where $T$ is temperature and $s$ is entropy density.  There are two independent
differential equations of motion to solve for the functions $T(r)$ and $v(r)$.
These may succinctly written as
\begin{eqnarray}
\frac{d}{dr} \left( r^2 T^{0r} \right) &=& 0 \nonumber \\
\frac{d}{dr} \left( r^2 T^{rr} \right) &=& 0 \, .
\end{eqnarray}

An integral form of these equations is sometimes more useful since it can 
readily incorporate the input luminosity from the black hole.  The first
represents the equality of the energy flux passing through a sphere of radius r
with the luminosity of the black hole.
\begin{equation}
4\pi r^2 T^{0r} = L
\end{equation}
The second follows from integrating a linear combination of the differential
equations.  It represents the combined effects of the entropy from the black
hole together with the increase of entropy due to viscosity.
\begin{equation}
4\pi r^2 u s  = 4\pi \int_{r_0}^r dr' \, r'^2 \frac{1}{T}\left[
\frac{8}{9} \eta \left( \frac{du}{dr'} - \frac{u}{r'} \right)^2
+ \zeta \left( \frac{du}{dr'} + \frac{2u}{r'} \right)^2 \right]
+ \frac{L}{T_H}
\end{equation}
The term $L/T_H$ arises from equating the entropy per unit time lost by the
black hole $-d S_{\rm bh}/dt$ with that flowing into the matter.  Using the area
formula for the entropy of a black hole, $S_{\rm bh} = m_{\rm P}^2 \pi r_S^2 =
4\pi M^2/m_{\rm P}^2$, and identifying $-dM/dt$ with the luminosity, the entropy
input from the black hole is obtained.

The above pair of equations are to be applied beginning at some radius $r_0$
greater than the Schwarzschild radius $r_S$, that is, outside the quantum
particle production region of the black hole.
The radius $r_0$ at which the imperfect fluid equations are first applied should
be chosen to be greater than the Schwarzschild radius, otherwise the computation
of particle creation by the black hole would be invalid.  It should not be too
much greater, otherwise particle collisions would create more entropy than is
accounted for by the equation above.  The energy and entropy flux into the fluid 
come from quantum particle creation by the black hole at temperature $T_H$. 
Gravitational effects are of order $r_S/r$, hence
negligible for $r > (5-10)r_S$.

\section{Equation of State and Transport Coefficients}

Determination of the equation of state as well as the two viscosities for
temperatures ranging from MeV to TeV and more is a formidable task.  Here we
shall present some relatively simple parametrizations that seem to contain the 
essential physics.  Improvements to these can certainly be made, but probably 
won't change the viscous fluid flow or the observational consequences very much. 

The hot shell of matter surrounding a primordial black hole provides a
theoretical testing ground rivaled only by the big bang itself. 
To illustrate this we have plotted a semi-realistic parametrization of the 
equation of state in figure 1.  Gravitons and neutrinos are not included.  We 
assume (for fun) a second order electroweak phase transition at a temperature of 
$T_{EW}$ = 100 GeV.  Above that temperature the standard model has 101.5 
effective massless bosonic degrees of freedom (as usual fermions count as 7/8 of 
a boson).  We assume (also for fun) a first order QCD phase transition at a 
temperature of $T_{QCD}$ = 170 MeV.  The number 
of effective massless bosonic degrees of freedom changes from 47.5 just above 
this critical temperature (u, d, s quarks and gluons) to 7.5 just below it 
(representing the effects of all the massive hadrons in the particle data 
tables) \cite{OK}.  Below 30 MeV only electrons, positrons, and photons remain, 
and finally below a few hundred keV only photons survive in any appreciable 
number.  The explicit parametrization shown in figure 1 is as follows.
\begin{eqnarray}
{s(T) \, = \, \frac{4\pi^2}{90}T^3 \,
\left\{ \begin{array}{ll}
101.5 & T_{EW} \leq T \\
56.5 + 45 \,{\rm e}^{-(T_{EW}-T)/T} & T_{QCD} \leq T < T_{EW} \\
2 + 3.5 \,{\rm e}^{-m_e/T} + 27.25 \,{\rm e}^{-(T_{QCD}-T)/T}
 & T < T_{QCD}
\end{array} \right. }
\end{eqnarray}
A word about neutrinos:  It is quite possible that they should be considered in 
approximate equilibrium at temperatures above 100 GeV where the electroweak 
symmetry is restored.  Still there is some uncertainty about this.  Since they 
provide only a few effective degrees of freedom out of more than 100 their 
neglect should cause negligible error.  We will address neutrinos and their 
emission in a subsequent paper.

Now we turn to the viscosities.  The shear viscosity was calculated in 
\cite{tau} for the full standard model in the symmetry restored phase, meaning 
temperatures above 100 GeV or so, using the relaxation time approximation.  The 
result is
\begin{equation}
\eta(T > 100 \, {\rm GeV}) = 82.5 \, T^3
\end{equation}
when numerical values for coupling constants etc. are put in.  The shear 
viscosity for QCD degrees of freedom only was calculated to leading order in the 
QCD coupling $\alpha_s$ in \cite{baym} to be
\begin{equation}
\eta({\rm QCD}) = \frac{0.342 (1+1.7N_f)}{(1+N_f/6)\alpha_s^2
\ln(\alpha_s^{-1})} \, T^3
\end{equation}
where $N_f$ is the number of quark flavors whose mass is less than $T$.  This is 
the extent of our knowledge of shear viscosity at high temperature.  We observe 
that the ratio of the shear viscosity to the entropy density, as appropriate for 
the above two cases, is dimensionless and has about the same numerical value in 
both.  Therefore, as a practical matter we assume that the shear viscosity 
always scales with the entropy density for all temperatures of interest.  We 
take the constant of proportionality from the full standard model cited above.
\begin{equation}
\eta = \frac{82.5}{101.5} \left( \frac{s}{4\pi^2/90} \right)
\end{equation}
There is even less known about the bulk viscosity at the temperatures of 
interest to us. The bulk viscosity is zero for point particles with no internal 
degrees of freedom and with local interactions among them.  In renormalizable 
quantum field theories the interactions are not strictly local.  In particular, 
the coupling constants acquire temperature dependence according to the 
renormalization group.  For example, to one loop order the QCD coupling has the 
functional dependence $\alpha_s \sim 1/\ln(T/\Lambda)$ where $\Lambda$ is the 
QCD scale.  On account of this dependence the bulk viscosity is nonzero.  We 
estimate that
\begin{equation}
\zeta \approx 10^{-4}\, \eta
\end{equation}
and this is what we shall use in the numerics.      

Overall we have a modestly realistic description of the equation of state and 
the viscosities that are still a matter of theoretical uncertainty.
One needs $s(T), \eta(T), \zeta(T)$ over a huge range of $T$.  Of course, these 
are some of the quantities one hopes to obtain experimental information on from 
observations of exploding black holes.

\section{Numerical Solution and Scaling}

Several limiting cases of the relativistic viscous fluid equations were
studied in \cite{me}.  The most realistic situation used the equation of
state $\epsilon = aT^4$, $s = (4/3)aT^3$ and viscosities
$\eta = b_ST^3$, $\zeta = b_BT^3$ with the coefficients $a$, $b_S$, $b_B$
all constant.  A scaling solution, valid at large radii when $\gamma \gg 1$,
was found to be $T(r) = T_0 (r_0/r)^{2/3}$ and $\gamma(r) = \gamma_0 
(r/r_0)^{1/3}$.  The constants must be related by $36aT_0r_0 = 
(32b_S + 441b_B)\gamma_0$.  This $r$-dependence of $T$ and $\gamma$ is exactly 
what was conjectured by Heckler \cite{Hb}.

It was shown in \cite{me} that if approximate local thermal equilibrium is 
achieved it can be maintained, at least for the semi-realistic situation 
described above.  The requirement is that the inverse of the local volume 
expansion rate $\theta = u^{\mu}_{\;\; ;\mu}$ be
comparable to or greater than the relaxation time for thermal equilibrium
\cite{MTW}.  Expressed in terms of a local volume element $V$ and proper time
$\tau$ it is $\theta = (1/V)dV/d\tau$, whereas in the rest frame of the black
hole the same quantity can be expressed as $(1/r^2)d(r^2 u)/dr$.  Explicitly
\begin{equation}
\theta = \frac{7\gamma_0}{3r_0}\left(\frac{r_0}{r}\right)^{2/3}
= \frac{7\gamma_0}{3r_0T_0} T \, .
\end{equation}
Of prime importance in achieving and maintaining local thermal equilibrium in a
relativistic plasma are multi-body processes such as $2 \rightarrow 3$ and
$3 \rightarrow 2$, etc.  This has been well-known when calculating quark-gluon
plasma formation and evolution in high energy heavy ion collisions
\cite{klaus,S}
and has been emphasized in ref. \cite{Ha,Hb} in the context of black hole
evaporation.  This is a formidable task in the standard model with its 16
species of particles.  Instead three estimates for the requirement that
local thermal equilibrium be maintained were made. The first and simplest 
estimate is to require that the thermal DeBroglie wavelength of a massless 
particle, $1/3T$, be less than $1/\theta$.   The second estimate is to require 
that the Debye screening length for each of the gauge groups in the standard 
model be less than $1/\theta$. The Debye screening length is the inverse of the 
Debye screening mass $m^{\rm D}_n$ where $n =1, 2, 3$ for the gauge groups U(1),
SU(2), SU(3).  Generically $m^{\rm D}_n \propto g_nT$ where $g_n$ is the gauge
coupling constant and the coefficient of proportionality is essentially the
square root of the number of charge carriers \cite{kapbook}.  For example, for
color SU(3) $m^{\rm D}_3 = g_3 \sqrt{1+N_{\rm f}/6}\,T$
where $N_{\rm f}$ is the number of light quark flavors at the temperature $T$.
The numerical values of the gauge couplings are: $g_1 = 0.344$, $g_2 = 0.637$,
and $g_3 = 1.18$ (evaluated at the scale $m_Z$) \cite{PDG}.  So within a factor
of about 2 we have $m^{\rm D} \approx T$. The third and most relevant estimate
is the mean time between two-body collisions in the standard model for
temperatures greater than the electroweak symmetry restoration temperature.
This mean time was calculated in \cite{tau} in the process of
calculating the viscosity in the relaxation time approximation.  Averaged over
all particle species in the standard model one may infer from that paper an
average time of $3.7/T$.  Taking into account multi-body reactions would
decrease that by about a factor of two to four.  All three of these estimates
are consistent within a factor of 2 or 3.  The conclusion to be drawn is that
local thermal equilibrium should be achieved when
$\theta \lord T$. Once thermal equilibrium is achieved it is not lost because 
$\theta/T$ is independent of $r$.  The picture that emerges is that of an 
imperfect fluid just marginally kept in local equilibrium by viscous forces.

The results quoted above are only valid at large $r$ and for the equation of 
state $s \propto T^3$.  To know the behavior of the solution at non-asymptotic 
$r$ and for the more sophisticated equation of state and viscosities described 
in section 5 requires a numerical analysis.  We have found that the most 
convenient form of the viscous fluid equations for numerical evaluation are
\begin{equation}
4\pi r^2 \left[ \gamma u T s -\frac{4}{3} \eta \gamma u
\left( \frac{du}{dr} - \frac{u}{r} \right) -
\zeta \gamma u \left( \frac{du}{dr} + \frac{2u}{r} \right)
\right] = L
\end{equation}
for energy conservation (from eq. (9)) and
\begin{equation}
\frac{d}{dr} \left( 4\pi r^2 u s \right) = \frac{4\pi r^2}{T}
\left[ \frac{8}{9} \eta \left( \frac{du}{dr} - \frac{u}{r} \right)^2
+ \zeta \left( \frac{du}{dr} + \frac{2u}{r} \right)^2 \right]
\end{equation}
for entropy flow (from eq. (10)).  Obviously the entropy flux is a monotonically 
increasing function of $r$ because of dissipation.

Mathematically the above pair of equations apply for all $r > 0$, although 
physically we should only apply them beyond the Schwarzschild radius $r_S$.
Let us study them first in the limit $r \rightarrow 0$, which really means the 
assumption that $v \ll 1$.  Then $u \approx v$ and $\gamma \approx 1$.  We also 
consider black hole temperatures greater than $T_{EW}$ so that the equation of 
state and the viscosities no longer change their functional forms.  It is 
straightforward to check that a power solution satisfies the equations, with
\begin{eqnarray}
u(r) &=& u_i (r/r_i)^{2/5} \nonumber \\
T(r) &=& T_i (r_i/r)^{3/5}
\end{eqnarray}
where $r_i$ is some reference radius.  If the luminosity and the reference 
radius are given then $u_i$ and $T_i$ are determined by the fluid equations.

The numerical solution for all radii needs some initial conditions.  Typically 
we begin the solution at one-tenth the Schwarzschild radius.  At this radius the 
$u_i$, as determined above, is small enough to serve as a good first estimate.  
However, it needs to be fine-tuned to give an acceptable solution at large $r$.  
For example, at large $r$ there is an approximate but false solution:
$T = constant$ with $u \sim r$.  The problem is that we need a solution valid 
over many orders of magnitude of $r$.  If eq. (17) is divided by
$r^2$ and if the right hand side is neglected in the limit
$r \rightarrow \infty$ then the left hand side is forced to be zero.
We have used both 
Mathematica and a fourth order Runge-Kutta method with adaptive step-size to 
solve the equations.  They give consistent results.

In figure 2 we plot $u(r_S/10)$ versus $T_H$.  It is essentially constant for 
all $T_H > T_{EW}$ with the value of 0.0415.  In figure 3 we plot the function
$u(r)$ versus $r$ for three different black hole temperatures.  The radial 
variable $r$ is expressed in units of its value when the temperature first 
reaches $T_{QCD}$, and $u$ is expressed in units of its value at that same 
radius.  This allows us to compare different black hole temperatures.  To rather 
good accuracy these curves seem to be universal as they essentially lie on top 
of one another.  The curves are terminated when the temperature reaches 10 MeV.  
The function $u(r)$ behaves like $r^{1/3}$ until temperatures of order 100 MeV 
are reached.  The simple parametrization
\begin{equation}
u(r) = u_S \left( r/r_S \right)^{1/3}
\end{equation}
with $u_S = 0.10$
will be very useful when studying radiation from the surface of the fluid.

In figure 4 we plot the temperature in units of $T_{QCD}$ versus the radius in 
units of $r_{QCD}$ for the same three black hole temperatures as in figure 3.  
Again the curves are terminated when the temperature drops to 10 MeV.  The 
curves almost fall on top of one another but not perfectly.  The temperature 
falls slightly slower than the power-law behavior $r^{-2/3}$ expected on the 
basis of the equation of state $s = (4/3)aT^3$.  The reason is that the 
effective number of degrees of freedom is falling with the temperature.  The 
entropy density is shown in figure 5.  It also exhibits an imperfect degree of 
scaling similar to the temperature.

Since viscosity plays such an important role in the outgoing fluid we should 
expect significant entropy production.  In figure 6 we plot the entropy flow 
$4\pi r^2 u s$ as a function of radius for the same three black hole 
temperatures as in figures 3-5.  It increases by several orders of magnitude.  
The fluid flow is far from isentropic.

\section{Onset of Free-Streaming}

Eventually the fluid expands so rapidly that the particles composing the fluid 
lose thermal contact with each other and begin free-streaming.  In heavy ion 
physics this is referred to as thermal freeze-out, and in astronomy it is 
usually associated with the photosphere of a star.  In 
the sections above we argued that thermal contact should occur for all 
particles, with the exception of gravitons and neutrinos, down to temperatures 
on the order of $T_{QCD}$.  Below that temperature the arguments given no longer 
apply directly; for example, the relevant interactions are not those of 
perturbative QCD.

Extensive studies have been made of the interactions among hadrons at finite 
temperature.  Prakash {\it et al.} \cite{Prakash} used experimental information 
to construct scattering amplitudes for pions, kaons and nucleons and from them 
computed thermal relaxation rates.  The relaxation time for $\pi - \pi$ 
scattering can be read off from their figures and simply paramterized as
\begin{equation}
\tau^{-1}_{\pi \pi} \approx 16 \left( \frac{T}{100 \, {\rm MeV}} \right)^4
\,\, {\rm MeV}
\end{equation}
which is valid for $100 < T < 200$ MeV.  This rate is compared to the volume
expansion rate $\theta$ (see section 5) in figure 7.  From the figure it is 
clear that pions cannot maintain thermal equilibrium much below 160 MeV or so.  
Since pions are the lightest hadrons and therefore the most abundant at low 
temperatures, it seems unlikely that other hadrons could maintain thermal 
equilibrium either.

Heckler has argued vigorously that electrons and photons should continue to 
interact down to temperatures on the order of the electron mass \cite{Ha,Hb}.  
Multi-particle reactions are crucial to this analysis.  Let us see how it 
applies to the present situation.  Consider, for example, the cross section for 
$ee \rightarrow ee\gamma$.  The energy-averaged cross section is \cite{Hb}
\begin{equation}
\overline{\sigma}_{\rm brem} = 8 \alpha_{EM} r_0^2 \ln(2E/m_e)
\end{equation}
where $m_e$ is the electron mass, $r_0 = \alpha_{EM}/m_e$ is the classical 
electron radius, and $E \gg m_e$ is the energy of the incoming electrons in the
center-of-momentum frame.  (If one computes the rate for a photon produced with 
the specific energies $0.1E$, $0.25E$, or $0.5E$ the cross section would be 
larger by a factor 4.73, 2.63, or 1.27, respectively.)  The rate using the 
energy-averaged cross section is
\begin{equation}
\tau^{-1}_{\rm brem} \approx \left[ \frac{3}{\pi^2} \xi(3) T^3 \right]
\left[\frac{8\alpha^3_{EM}}{m_e^2} \ln (6T/m_e) \right]
\end{equation}
where we have used the average energy $\langle E \rangle \approx 3T$ for 
electrons with $m_e \ll T$.  This rate is also plotted in figure 7.  It is large 
enough to maintain local thermal equilibrium down to temperatures on the order 
of 140 MeV.  Of course, there are other electromagnetic many-particle reactions 
which would increase the overall rate.  On the other hand, as pointed out by 
Heckler \cite{Hb}, these reactions are occurring in a high density plasma with 
the consequence that dispersion relations and interactions are renormalized by 
the medium.  If one takes into account only renormalization of the electron 
mass, such that $m_{eff}^2 \approx m_e^2 + e^2 T^2/3$ when $m_e \ll T$, then the 
rate would be greatly reduced.

Does this mean that photons and electrons are not in thermal equilibrium at the 
temperatures we have been discussing?  Consider bremsstrahlung reactions in the 
QCD plasma.  There are many $2 \rightarrow 3$ reactions, such as: $q_1 q_2
\rightarrow q_1 q_2 \gamma$, $q_1 \overline{q}_2 \rightarrow q_1 \overline{q}_2
\gamma$, $gg \rightarrow q \overline{q} \gamma$, and so on.  Here the subscripts 
label the quark flavor, which may or may not be the same.  The rate for these 
can be estimated using known QED and QCD cross sections 
\cite{Jauch,Haug,Cutler}.  Using an effective quark mass given by $gT$ we find 
that the rate is $\alpha_s T$ with a coefficient of order or larger than unity.  
Since $\alpha_s$ becomes of order unity near $T_{QCD}$ we conclude that photons 
are in equilibrium down to temperatures of that order at least.  To make the 
matter even more complicated we must remember that the expansion rate $\theta$ 
is based on a numerical solution of the viscous fluid equations which assume a 
constant proportionality between the shear and bulk viscosities and the entropy 
density.  Although these proportionalities may be reasonable in QCD and 
electroweak plasmas at high temperatures they may fail at temperatures below 
$T_{QCD}$.  The viscosities should be computed using the relaxation times for 
self-consistency of the transition from viscous fluid flow to free-streaming, 
which we have not done.  For example, the first estimate for the shear viscosity 
for massless particles with short range interactions is $T^4 \tau$ where $\tau$ 
is the relaxation time.  For pions we would get $\eta \sim constant$, not
$\eta \sim T^3$.  As another example, we must realize that the bulk viscosity 
can become significant when the particles can be excited internally.  This is, 
in fact, the case for hadrons.  Pions, kaons and nucleons are all the lowest 
mass hadrons each of which sits at the base of a tower of resonances \cite{PDG}.
See, for example, \cite{Weinberg} and references therein.

In order to do gamma ray phenomenology we need a practical criterion for the 
onset of free-streaming.  We shall assume that this happens suddenly at a 
temperature $T_f$ in the range 100 to 140 MeV.  We shall assume that particles 
whose mass is significantly greater than $T_f$ have all annihilated, leaving 
only photons, electrons, muons and pions.  In figure 8 we plot the freeze-out 
radius $r_f = r(T_f)$ for $T_f =$ 100, 120 and 140 MeV versus the Schwarzschild 
radius.  The fact that $r_f$ increases as $T_f$ decreases is an obvious 
consequence of energy conservation.  More interesting is the power-law scaling:
$r_f \sim r_S^{-1/2} \sim T_H^{1/2}$.  This scaling can be understood as 
follows.

The luminosity from the decoupling or freeze-out surface is
\begin{equation}
L_f = 4\pi r_f^2 \left( \frac{2\pi^2}{45} \gamma_f^2 T_f^4 \right) d_f
\end{equation}
where the quantity in parentheses is the surface flux for one massless bosonic 
degree of freedom and $d_f$ is the total number of effective massless bosonic 
degrees of freedom.  For the particles listed above we have $d_f = 12$.  By 
energy conservation this is to be equated with the Hawking formula for the black 
hole luminosity,
\begin{equation}
L_h = 64\pi^2 \alpha_h^{eff} T_H^2 \, ,
\end{equation}
where $\alpha_h^{eff}$ does not include the contribution from gravitons and 
neutrinos.  Together with the scaling function for the flow velocity, eq. (20), 
we can solve for the radius
\begin{equation}
r_f = \frac{2}{\pi} \left( \frac{45 \pi \alpha_h^{eff}}{2 u_S^2 d_f} 
\right)^{3/8} \sqrt{\frac{T_H}{T_f^3}}
\end{equation}
and for the boost
\begin{equation}
\gamma_f T_f = 2u_S \left( \frac{45 \pi \alpha_h^{eff}}{2 u_S^2 d_f} 
\right)^{1/8} \sqrt{T_f T_H} \approx 0.22 \sqrt{T_f T_H} \, .
\end{equation}
From these we see that the final radius does indeed scale like the inverse of 
the square-root of the Schwarzschild radius or like the square-root of the black 
hole temperature, and that the average particle energy (proportional to
$\gamma_f T_f$) scales like the square-root of the black hole temperature.  One 
important observational effect is that the average energy of the outgoing 
particles is reduced but their number is increased compared to direct Hawking 
emission into vacuum \cite{Ha,Hb}.

\section{Photon Emission}

Photons emitted into the vacuum surrounding the black hole primarily come from 
one of two sources.  Either they are emitted directly in the form of a boosted 
black-body spectrum, or they arise from neutral pion decay.  We will consider 
each of these in turn.

\subsection{Direct photons}

Photons emitted directly have a Planck distribution in the local rest frame
of the fluid.  The phase space density is
\begin{equation}
f(E') = \frac{1}{{\rm e}^{E'/T_f} -1} \, .
\end{equation}
The energy appearing here is related to the energy as measured in the rest frame 
of the black hole and to the angle of emission relative to the radial vector by
\begin{equation}
E' = \gamma_f (1-v_f \cos\theta) E \, .
\end{equation}
No photons will emerge if the angle is greater than $\pi/2$.  Therefore the 
instantaneous distribution is
\begin{eqnarray}
\frac{d^2N_{\gamma}^{\rm dir}}{dE dt} &=& 
4\pi r_f^2 \left(\frac{E^2}{2\pi^2}\right)
\int_0^1 d(\cos\theta) \cos\theta f(E,\cos\theta) \nonumber \\ & \approx &
- \frac{2 r_f^2 T_f E}{\pi \gamma_f} \ln \left(1 - 
{\rm e}^{-E/2 \gamma_f T_f} \right)
\end{eqnarray}
where the second equality holds in the limit $\gamma_f \gg 1$.  This limit is 
actually well satisfied for us and is used henceforth.

The instantaneous spectrum can be integrated over the remaining lifetime of the 
black hole straightforwardly.  The radius and boost are both known in terms of 
the Hawking temperature $T_H$, and the time evolution of the latter is simply 
obtained from solving eq.(3).  For a black hole that disappears at time $t = 0$ 
we have
\begin{equation}
T_H(t) = - \frac{1}{8\pi} \left( \frac{m_P^2}{3 \alpha_h t} \right)^{1/3} \, .
\end{equation}
Starting with a black hole whose temperature is $T_0$ we obtain the spectrum
\begin{equation}
\frac{dN_{\gamma}^{\rm dir}}{dE} = \frac{360 u_S^2}{\pi^5 d_f}
\left( \frac{45 \pi \alpha_h^{eff}}{2 u_S^2 d_f} \right)^{1/4}
\frac{m_P^2 T_f}{E^4}
\sum_{n=1}^{\infty} \frac{1}{n} \int_0^{E/2 \gamma_f(T_0) T_f}
dx \, x^4 \, {\rm e}^{-nx} \, .
\end{equation}
Here we have ignored the small numerical difference between $\alpha_h^{eff}$
and $\alpha_h$. In the high energy limit, namely, when $E \gg  2\gamma_f(T_0) 
T_f$, the summation yields the pure number $4(2\pi^6/315)$.  Note the power-law 
behavior $E^{-4}$.  This has important observational consequences.

\subsection{$\pi^0$ decay photons}

The neutral pion decays almost entirely into two photons: $\pi^0
\rightarrow \gamma \gamma$.  In the rest frame of the pion the single photon 
Lorentz invariant distribution is
\begin{equation}
E \frac{d^3N_{\gamma}}{d^3p} = \frac{\delta(E-m_{\pi}/e)}{\pi m_{\pi}}
\end{equation}
which is normalized to 2.  This must be folded with the distribution of $\pi^0$ 
to obtain the total invariant photon distribution.
\begin{equation}
E \frac{d^4N_{\gamma}}{d^3p dt} = \int_{m_{\pi}}^{\infty} dE_{\pi}
\frac{d^2N_{\pi^0}}{dE_{\pi}dt} \frac{1}{\pi m_{\pi}}
\delta \left( \frac{ EE_{\pi} - {\bf p} \cdot {\bf p}_{\pi} }
{m_{\pi}} - \frac{m_{\pi}}{2} \right)
\end{equation}
After integrating over angles we get
\begin{equation}
\frac{d^2N_{\gamma}^{\pi^0}}{dE dt} = 2 \int_{E_{\rm min}}^{\infty}
\frac{dE_{\pi}}{p_{\pi}} \frac{d^2N_{\pi^0}}{dE_{\pi}dt}
\end{equation}
where $E_{\rm min} = (E^2 + m_{\pi}^2/4)/E$.  In the limit $E \gg m_{\pi}$ we 
can approximate $E_{\rm min} = E$ and evaluate $d^2N_{\pi^0}/dE_{\pi}dt$ in 
the same way as photons.  This leads to the relatively simple expression
\begin{equation}
\frac{d^2N_{\gamma}^{\pi^0}}{dE dt} = \frac{4 r_f^2 T_f^2}{\pi^2} 
\sum_{n=1}^{\infty} \frac{1}{n^2} {\rm e}^{-nE/2\gamma_f T_f} \, .
\end{equation}
The time-integrated spectrum is computed in the same way as for direct photons.
\begin{equation}
\frac{dN_{\gamma}^{\pi^0}}{dE} = \frac{360 u_S^2}{\pi^5 d_f}
\left( \frac{45 \pi \alpha_h^{eff}}{2 u_S^2 d_f} \right)^{1/4}
\frac{m_P^2 T_f}{E^4}
\sum_{n=1}^{\infty} \frac{1}{n^2} \int_0^{E/2 \gamma_f(T_0) T_f}
dx \, x^3 \, {\rm e}^{-nx}
\end{equation}
In the high energy limit, namely, when $E \gg  2\gamma_f(T_0) T_f$, the 
summation yields the pure number $2\pi^6/315$.

\subsection{Instantaneous and integrated photon spectra}

The instantaneous spectra of high energy gamma rays, arising from both direct 
emission and from $\pi^0$ decay, are plotted in figures 9 (for $T_f = 140$ MeV) 
and 10 (for $T_f = 100$ MeV).  In each figure there are three curves 
corresponding to Hawking temperatures of 100 GeV, 1 TeV and 10 TeV.  
The photon spectra are essentially exponential above a few GeV with inverse 
slope $2 \gamma_f(T_H) T_f \propto \sqrt{T_f T_H}$.  If these instantaneous 
spectra could be measured they would tell us a lot about the equation of state, 
the viscosities, and how energy is processed from first Hawking radiation to 
final observed gamma rays.  Even the time evolution of the black hole luminosity 
and temperature could be inferred.

The time integrated spectra for $T_f = 140$ MeV are plotted in figure 11 for
three initial temperatures $T_0$.  A black hole with a Hawking temperature of 
100 GeV has 5.4 days to live, a black hole with a Hawking temperature of 1 TeV 
has 7.7 minutes to live, and a black hole with a Hawking temperature of 10 TeV 
has only 1/2 second to live.  The high energy gamma ray spectra are represented 
by
\begin{equation}
\frac{dN}{dE} = \frac{m_P^2 T_f}{26 E^4} \, .
\end{equation}
It is interesting that the contribution from $\pi^0$ decay comprises 20\% of the 
total while direct photons contribute the remaining 80\%.  The $E^{-4}$ fall-off 
is the same as that obtained by Heckler \cite{Ha}, whereas Halzen {\it et al.} 
\cite{Hal} and MacGibbon and Carr \cite{MC} obtained an $E^{-3}$ fall-off on the 
basis of direct fragmentation of quarks and gluons with no fluid flow and no 
photosphere.

\section{Observability of Gamma Rays}

The most obvious way to observe the explosion of a microscopic black hole is by 
high energy gamma rays.  We consider their contribution to the diffuse gamma ray 
spectrum in subsection 8.1, and in subsection 8.2 we study the systematics of a 
single identifiable explosion.

\subsection{Diffuse spectra from the galactic halo}

Suppose that microscopic black holes were distributed about our galaxy in some 
fashion.  Unless we were fortunate enough to be close to one so that we could 
observe its demise, we would have to rely on their contribution to the diffuse 
background spectrum of high energy gamma rays.

The flux of photons with energy greater than 1 GeV at Earth can be computed from 
the results of section 7 together with the knowledge of the rate density 
$\dot{\rho}({\bf x})$ of exploding black holes.  It is
\begin{equation}
\frac{d^3N_{\rm Earth}}{dE dA dt} = \frac{m_P^2 T_f}{26 E^4}
\int d^3x \frac{\dot{\rho}({\bf x})}{4\pi d^2({\bf x})}
{\rm e}^{-d({\bf x})/\lambda_{\gamma \gamma}(E)}
\end{equation}
where $d({\bf x})$ is the distance from the black hole to the Earth.  The 
exponential decay is due to absorption of the gamma ray by the black-body 
radiation \cite{Gould}.  The mean free path $\lambda_{\gamma \gamma}(E)$ is 
highly energy dependent.  It has a minimum of about 1 kpc around 1 PeV, and is 
greater than $10^5$ kpc for energies less than 100 TeV.

We need a model for the rate density of exploding black holes.  We shall assume 
they are distributed in the same way as the matter comprising the halo of 
our galaxy.  Thus we take
\begin{equation}
\dot{\rho}({\bf x}) = \dot{\rho}_0 \,
\frac{R_c^2}{x^2+y^2 +q^2 z^2 + R_c^2}
\end{equation}
where the galactic plane is the $x-y$ plane, $R_c$ is the core radius, and $q$ 
is a flattening parameter.  For numerical calculations we shall take the core 
radius to be 10 kpc.  The Earth is located a distance $R_E = 8.5$ kpc from the 
center of the galaxy and lies in the galactic plane.  Therefore
$d^2 = (x-R_E)^2 + y^2 + z^2$.

The last remaining quantity is the normalization of the rate density 
$\dot{\rho}_0$.  This is, of course, unknown since no one has ever knowingly 
observed a black hole explosion.  The first observational limit was determined 
by Page and Hawking \cite{PH}.  They found that the local rate density
$\dot{\rho}_{\rm local}$ is less than 1 to 10 per cubic parsec per year on the 
basis of diffuse gamma rays with energies on the order of 100 MeV.  This limit 
has not been lowered very much during the intervening twenty-five years.  For 
example, Wright \cite{W} used EGRET data to search for an anisotropic
high-lattitude component of diffuse gamma rays in the energy range from 30 MeV 
to 100 GeV as a signal for steady emission of microscopic black holes.  He 
concluded that $\dot{\rho}_{\rm local}$ is less than about 0.4 per cubic parsec 
per year.  (For an alternative point of view on the data see \cite{DBCline}.)
In our numerical calculations we shall assume a value
$\dot{\rho}_0 = 1$ pc$^{-3}$ yr$^{-1}$ corresponding to
$\dot{\rho}_{\rm local} \approx 0.58$ pc$^{-3}$ yr$^{-1}$.  This makes for easy 
scaling.  Estimating the quantity of dark matter in our galaxy as
$M_{\rm halo}/\rho_{0, \, {\rm halo}} = 4.7\times 10^4$ kpc$^3$ means that we 
could have up to $47\times 10^{12}$ microscopic black hole explosions per year 
in our galaxy.

Figure 12 shows the calculated flux at Earth, multiplied by $E^4$.  Of course 
this curve would be flat if it weren't for absorption on the microwave 
background radiation.  There is a relative suppression of three orders of 
magnitude centered between $10^{15}$ and $10^{16}$ eV.  This means that it is 
unlikely to observe exploding black holes in the gamma ray spectrum above 
$10^{14}$ eV.  Even below that energy it is unlikely because they have not been 
observed at energies on the order of 100 MeV, and the spectrum falls 
faster than the primary cosmic ray spectrum $\propto E^{2.7}$.  The curve 
displayed in figure 12 assumes a spherical halo, $q = 1$, but there is hardly 
any difference when the halo is flattened to $q = 2$.

\subsection{Point source systematics}

Given the unfavorable situation for observing the effects of exploding 
microscopic black holes on the diffuse gamma ray spectrum, we now turn to the 
consequences for observing one directly. How far away could one be seen?  Let us 
call that distance $d_{\rm max}$.  We assume that
$d_{\rm max} < \lambda_{\gamma\gamma}$ for simplicity, although that assumption 
can be relaxed if necessary.  Let $A_{\rm det}$ denote the effective area of the 
detector that can measure gamma rays with energies equal to or greater than 
$E_{\rm min}$.  The average number of gamma rays detected from a single 
explosion a distance $d_{\rm max}$ away is
\begin{equation}
\langle N_{\gamma}(E > E_{\rm min}) \rangle =
\frac{ A_{\rm det} }{4\pi d_{\rm max}^2} \int_{E_{\rm min}}^{\infty}
\frac{d N_{\gamma}}{dE} dE = \frac{ A_{\rm det} }{4\pi d_{\rm max}^2}
\frac{m_P^2 T_f}{78 E_{\rm min}^3} \, .
\end{equation}
Obviously we should have $E_{\rm min}$ as small as possible to get the largest
number, but it cannot be so small that the simple $E^{-4}$ behavior of the
emission spectrum is invalid.  See figure 11.

A rough approximation to the number distribution of detected gamma rays is a 
Poisson distribution.
\begin{equation}
P(N_{\gamma}) = \frac{\langle N_{\gamma} \rangle ^{N_{\gamma}}}
{N_{\gamma} !} {\rm e}^{-\langle N_{\gamma} \rangle }
\end{equation}
The exact form of the number distribution is not so important.  What is 
important is that when $\langle N_{\gamma}(E > E_{\rm min}) \rangle > 1$ we 
should expect to see multiple gamma rays coming from the same point in 
the sky.  Labeling these gamma rays according to the order in which they arrive, 
1, 2, 3, etc. we would expect their energies to increase with time:
$E_1 < E_2 < E_3 < ...$.  Such an observation would be remarkable, possibly 
unique, because astrophysical sources normally cool at late times.  This would 
directly reflect the increasing Hawking temperature as the black hole explodes 
and disappears.

It is interesting to know how the average gamma ray energy increases with time.
Using eqs. (30) and (36) we compute the average energy of direct photons to be
$4 \gamma_f T_f \zeta(4)/\zeta(3)$ and the average energy of $\pi^0$ decay 
photons to be one-half that.  The ratio of direct to decay photons turns out
to be $\pi$.  Therefore the average gamma ray energy is $3.17 \gamma_f(t) T_f$.
This average is plotted in figure 13 for $10^5 > t > 10^{-5}$ seconds.  The 
average gamma ray energy ranges from about 4 to 160 GeV. 

The maximum distance can now be computed.  Using some characteristic numbers we 
find
\begin{equation}
d_{\rm max} \approx 150 \sqrt{\frac{A_{\rm det}}{1 \, {\rm km}^2}}
\left( \frac{10 \, {\rm GeV}}{E_{\rm min}}\right)^{3/2} \, \, {\rm pc} \, .
\end{equation}
If we take the local rate density of explosions to be 0.4 pc$^{-3}$ yr$^{-1}$
then within 150 pc of Earth there would be $5\times 10^6$ explosions per year. 
These would be distributed isotropically in the sky.  Still, it suggests that 
the direct observation of exploding black holes is feasible if they are near to 
the inferred upper limit to their abundance in our neighborhood.  We should 
point out that a search for 1 s bursts of ultrahigh energy gamma rays from point 
sources by CYGNUS has placed an upper limit of $8.5 \times 10^5$ pc$^{-3}$
yr$^{-1}$ \cite{CYGNUS}.  However, as we have seen in figure 13 and
elsewhere, this is not what should
be expected if our calculations bear any resemblance to reality.
Rather than a burst, the luminosity and average gamma ray energy increase
monotonically over a long period of time.

\section{Conclusion}

The increasing energy of the radiated photons by an exploding black hole and
the disappearance of such a point gamma ray source in a certain period of time 
are unique characteristics of exploding black holes that may help us to detect 
them.  Still, there is much work to be done in determining whether the 
matter surrounding a black hole can reach and maintain thermal equilibrium.  The 
equation of state should be improved, and the viscosities computed using the 
relaxation times for self-consistency of the transition from viscous fluid flow 
to free streaming.  Also, there should be a more fundamental investigation of the relaxation times starting from the microscopic interactions.

Our next step is to calculate the neutrino flux radiated by exploding black 
holes.  Neutrino cross sections become very small at energies below 100 Gev 
which is the temperature of the electroweak phase transition.  Above 100 GeV the 
neutrino cross sections are the same as other particles in the standard model 
which allows neutrinos to interact enough to reach thermal equilibrium.  
Therefore neutrinos are expected to freeze out at a temperature around 100 GeV.  
Another worthwhile project is to carry out cascade simulations of the 
spherically expanding matter around the exploding black hole at a level of 
sophistication comparable to that of high energy heavy ion collisions. This 
project is much more complicated than the cascade simulation in heavy ion 
collision, though, because we need to deal with a much wider range of energies 
and particles involved in exploding black holes.  

The study of primordial black holes might well lead to great 
advancements in fundamental physics.  Because the highest temperatures in 
the universe exist in primordial black holes, matter at extremely high 
temperatures can be studied, and physics beyond the standard model can be 
tested.  In addition, because it is believed that baryon number is violated at 
high temperatures, the study of primordial black holes could possibly answer the 
question of why our universe became matter-dominated.  Because primordial black 
holes explode, they are an ideal model for studying the Big Bang and the birth 
of our universe.  Finally, the study of primordial black holes will 
help us to determine whether they are the source of the highest 
energy cosmic rays.  The origin of these cosmic rays is still one of the 
biggest mysteries today. 
Observation and experimental detection of exploding black holes will be one 
of the great challenges in the new millennium.  

\section*{Acknowledgements}

This work was supported by the US Department of Energy under grant
DE-FG02-87ER40328.

\newpage

\begin{figure}[h]
\centerline{\epsfig{figure=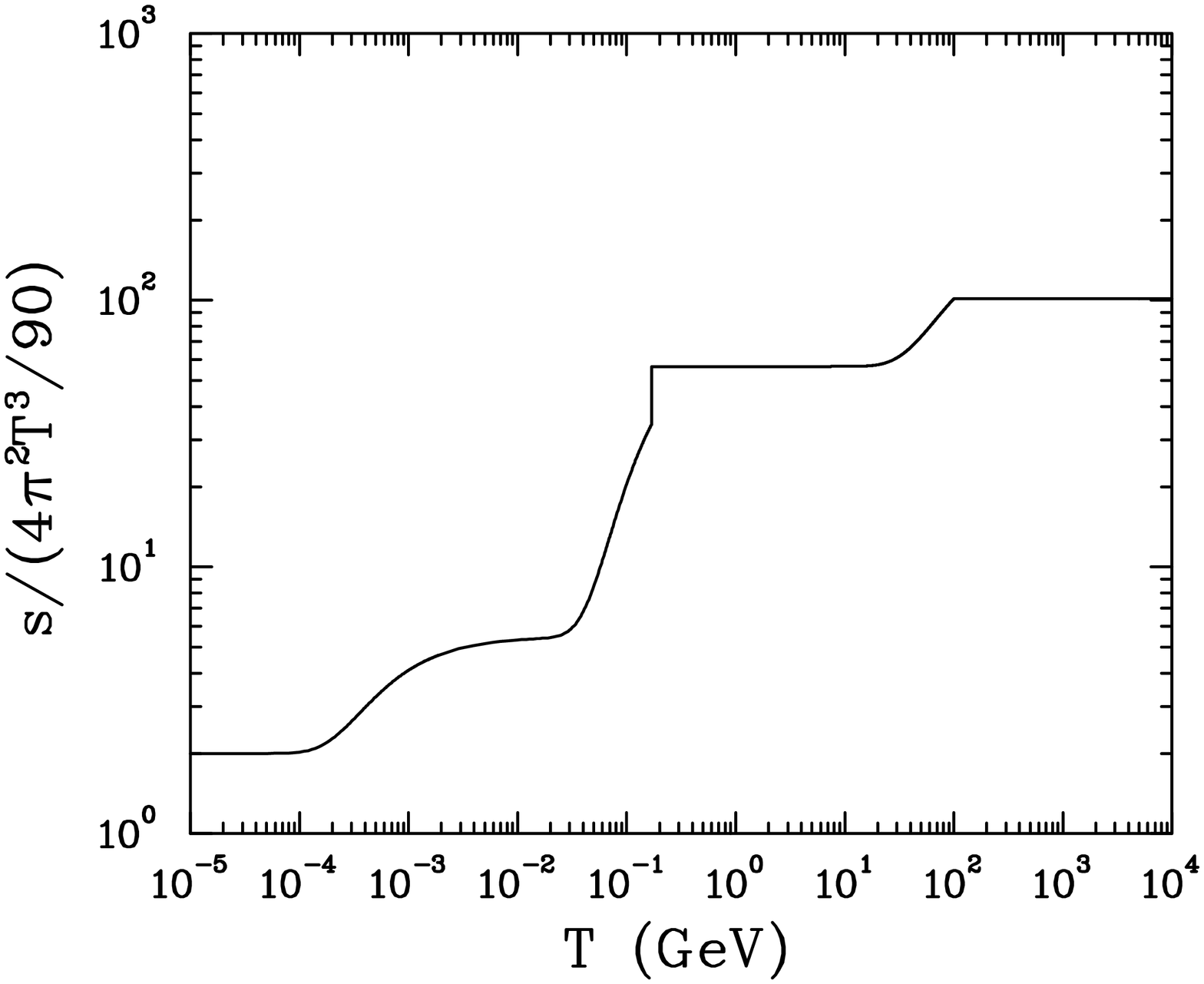,width=12.0cm,angle=90}}
\caption{Entropy density as a function of temperature, excluding neutrinos
and gravitons.  It is assumed that the QCD phase transition is first order
and the EW phase transition is second order.}
\end{figure}

\begin{figure}
\centerline{\epsfig{figure=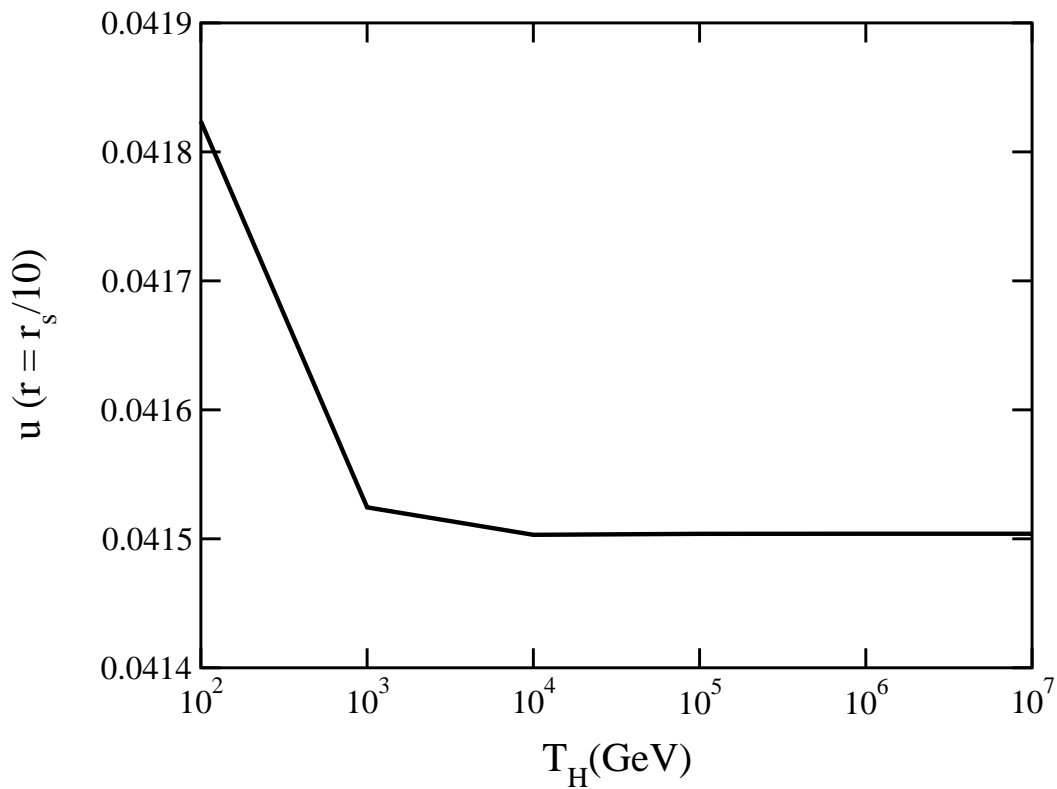,width=13.0cm,angle=270}}
\caption{The value of $u = v\gamma$ at one-tenth the Schwarzschild radius
as determined by numerical solution.  The physical applicability of the
numerical solution begins at radii greater than $r_S$.}
\end{figure}

\begin{figure}
\centerline{\epsfig{figure=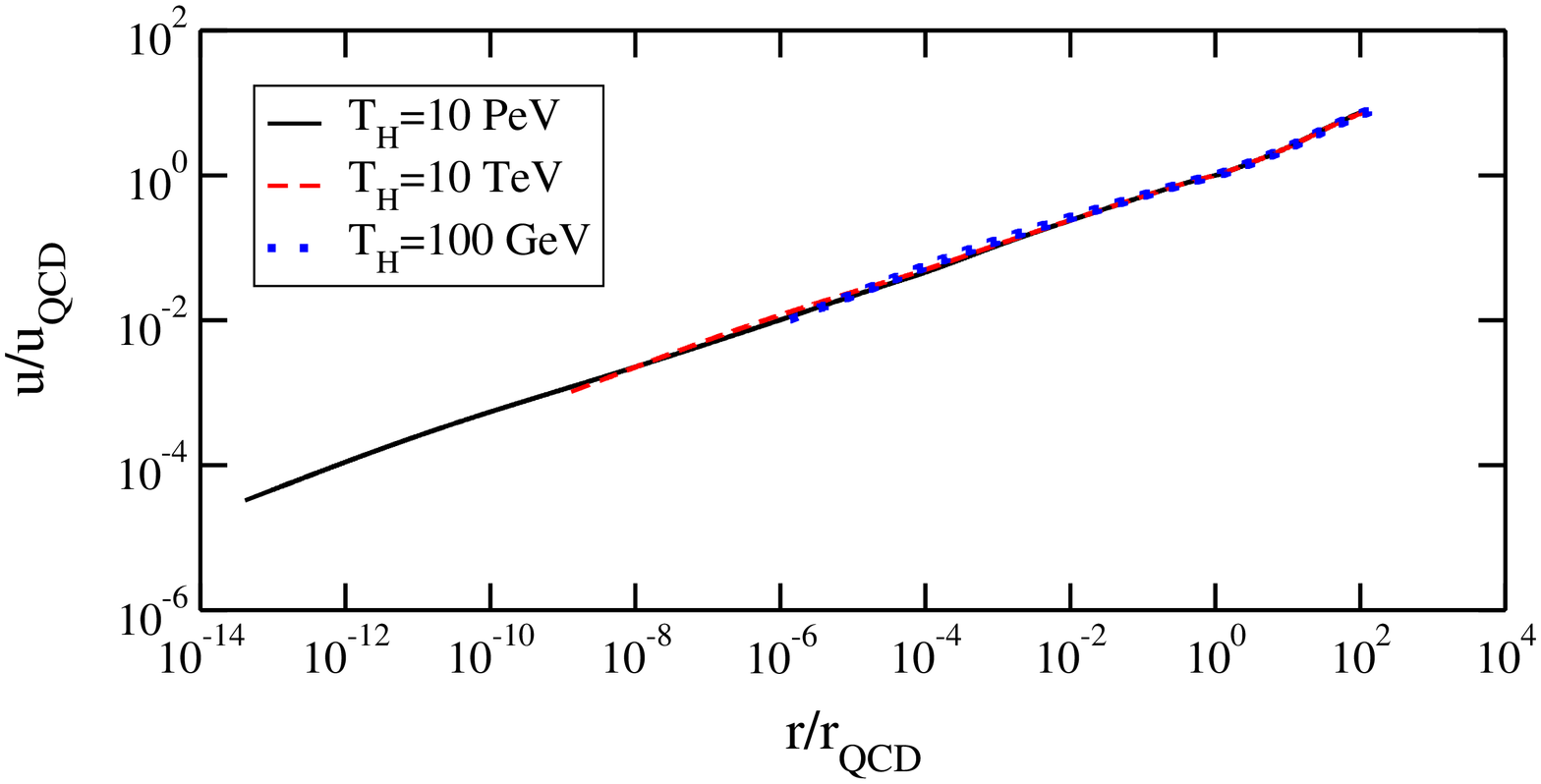,width=13.0cm,angle=270}}
\caption{The radial dependence of $u$ for three different Hawking temperatures.
The curves begin at $r_S/10$ and terminate when the local temperature reaches
10 MeV.}
\end{figure}

\begin{figure}
\centerline{\epsfig{figure=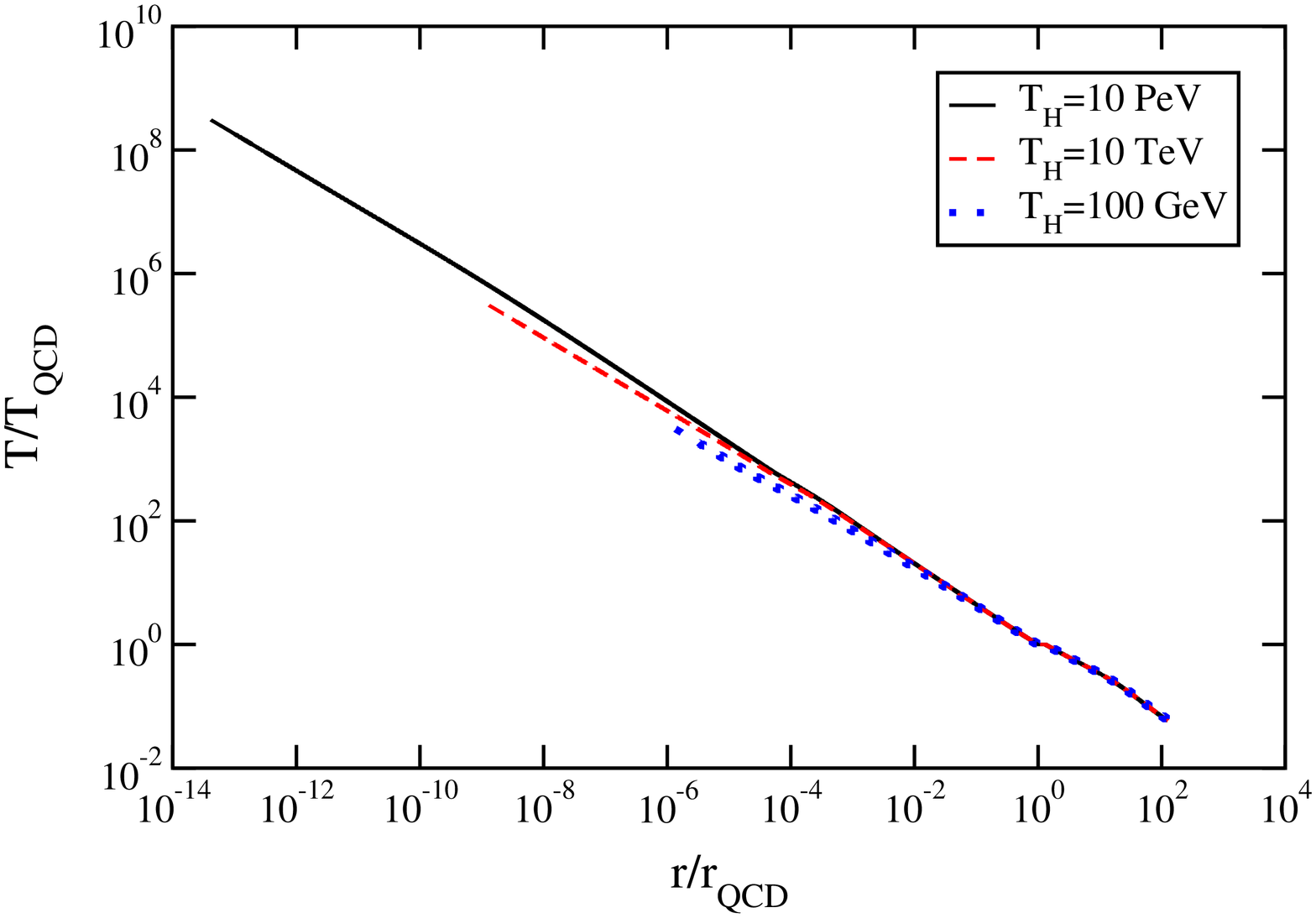,width=13.0cm,angle=270}}
\caption{ The radial dependence of $T$ for three different Hawking temperatures.
The curves begin at $r_S/10$ and terminate when the local temperature reaches
10 MeV.}
\end{figure}

\begin{figure}
\centerline{\epsfig{figure=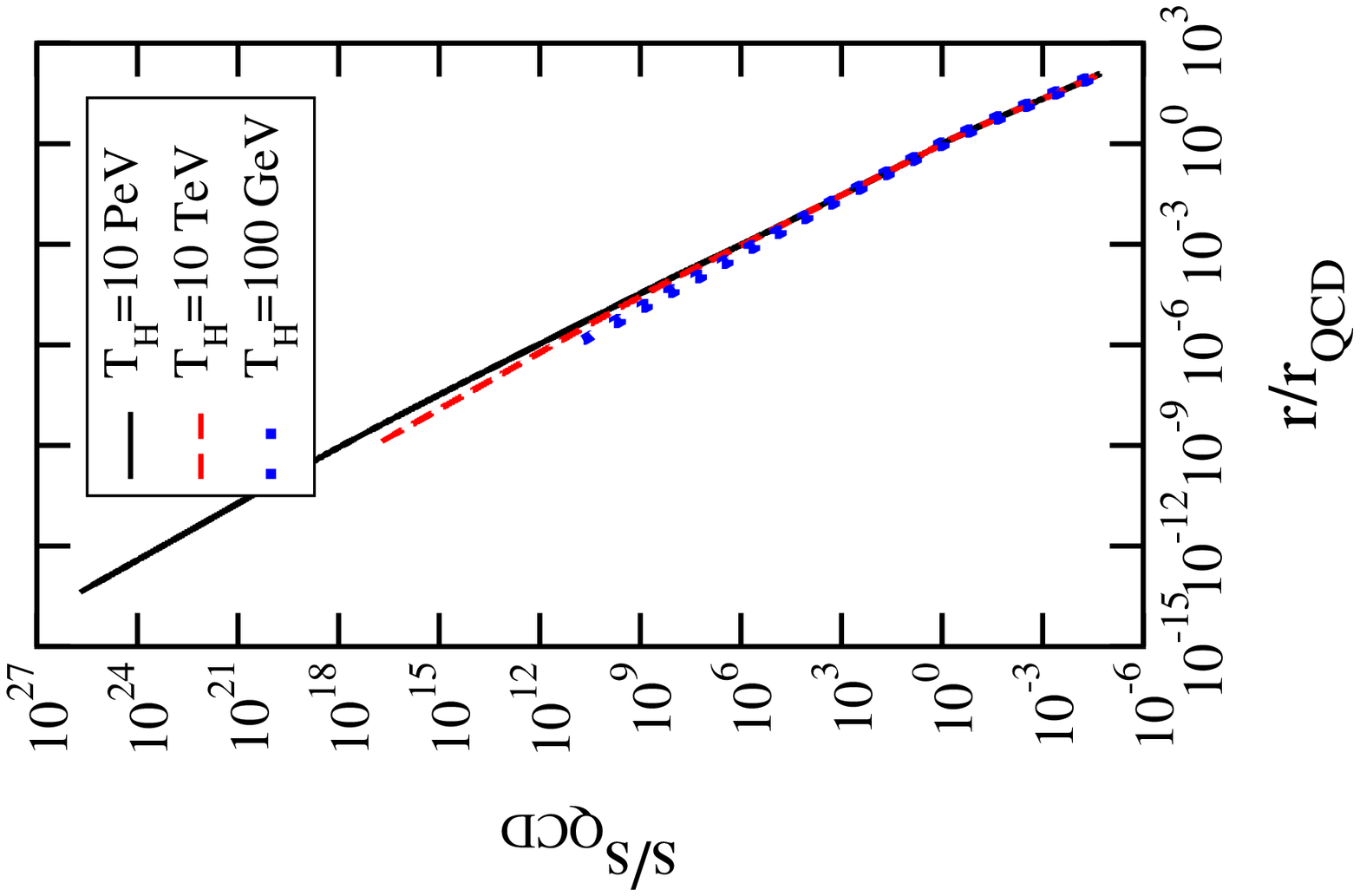,width=20.0cm,angle=270}}
\caption{The radial dependence of $s$ for three different Hawking temperatures.
The curves begin at $r_S/10$ and terminate when the local temperature reaches
10 MeV.}
\end{figure}

\begin{figure}
\centerline{\epsfig{figure=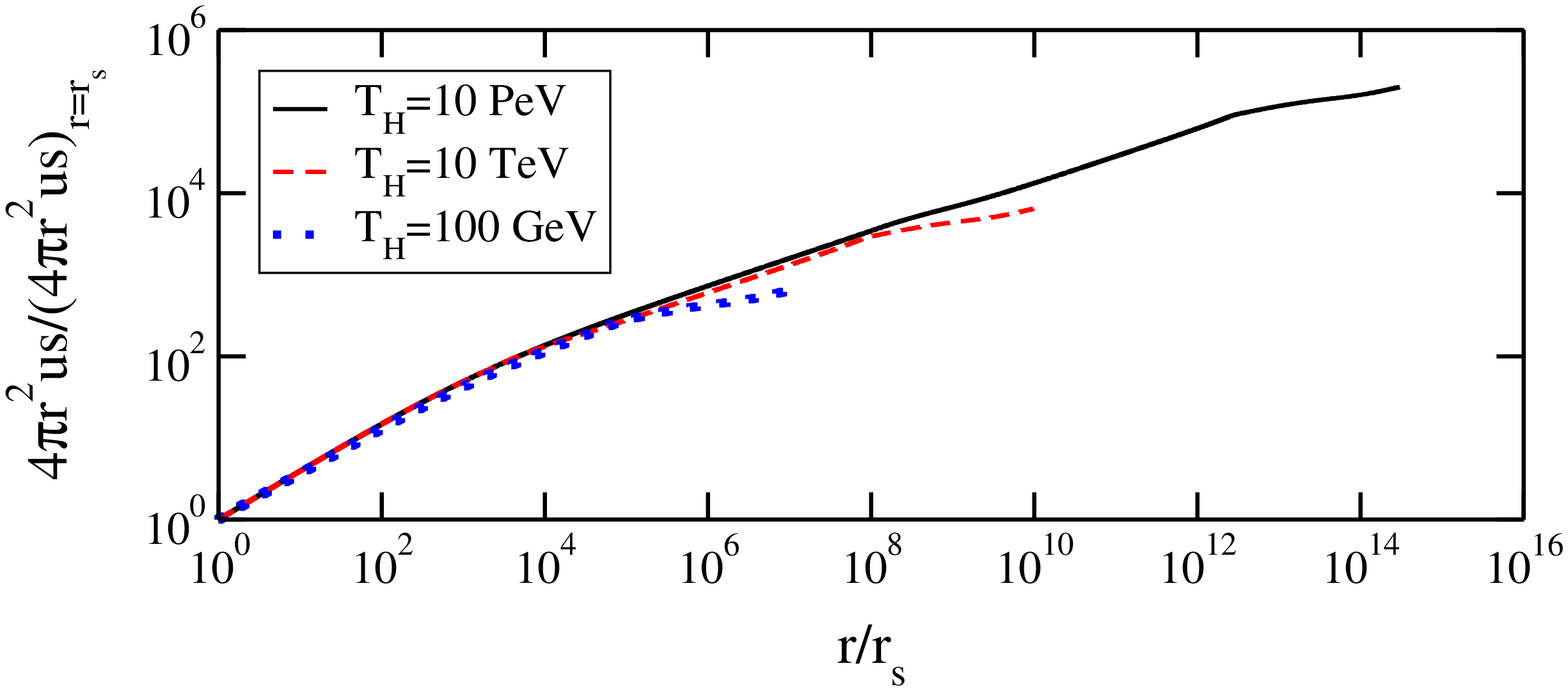,width=13.0cm,angle=270}}
\caption{The radial dependence of the entropy flow for three different Hawking 
temperatures.  The curves begin at $r_S/10$ and terminate when the local 
temperature reaches 10 MeV.}
\end{figure}

\begin{figure}
\centerline{\epsfig{figure=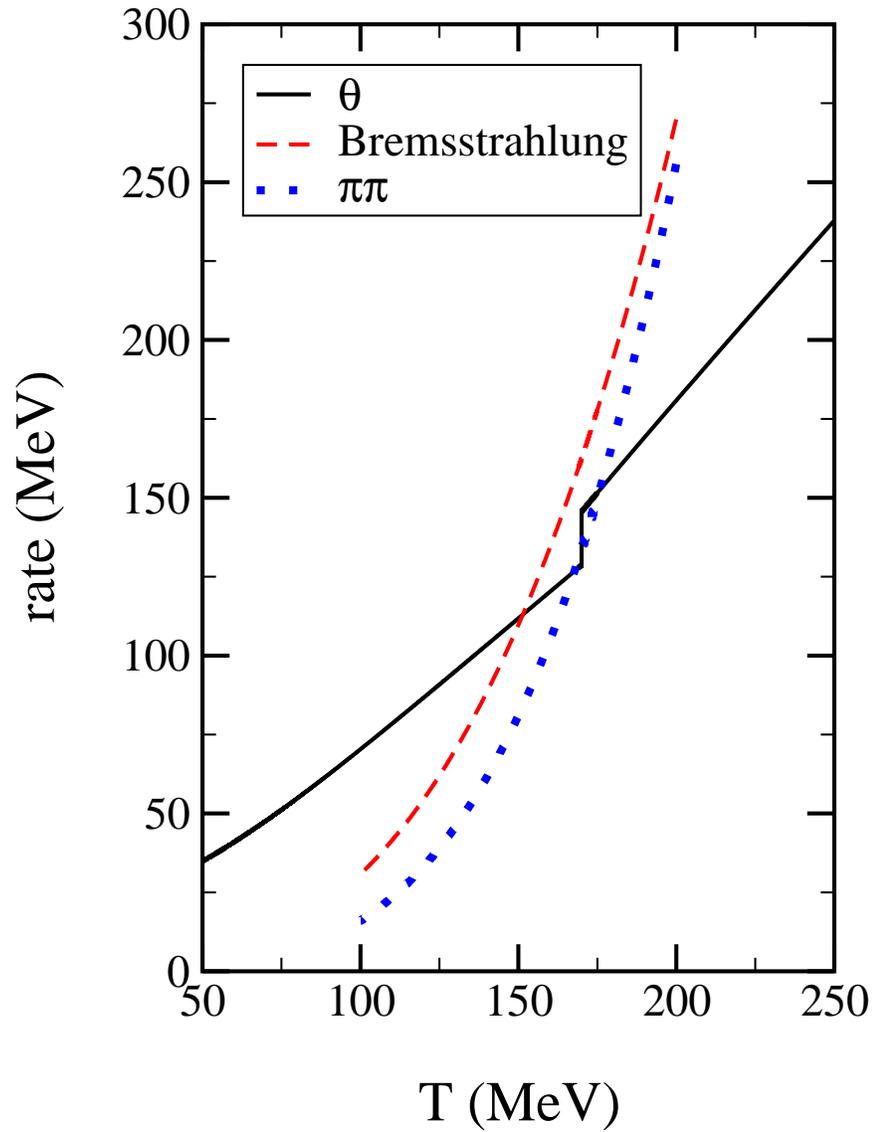,width=18.0cm,angle=270}}
\caption{The rate for $\pi\pi$ scattering and for the bremsstrahlung reaction
$ee \rightarrow ee\gamma$ are compared to the local volume expansion rate.
The Hawking temperature is 10 TeV.}
\end{figure}

\begin{figure}
\centerline{\epsfig{figure=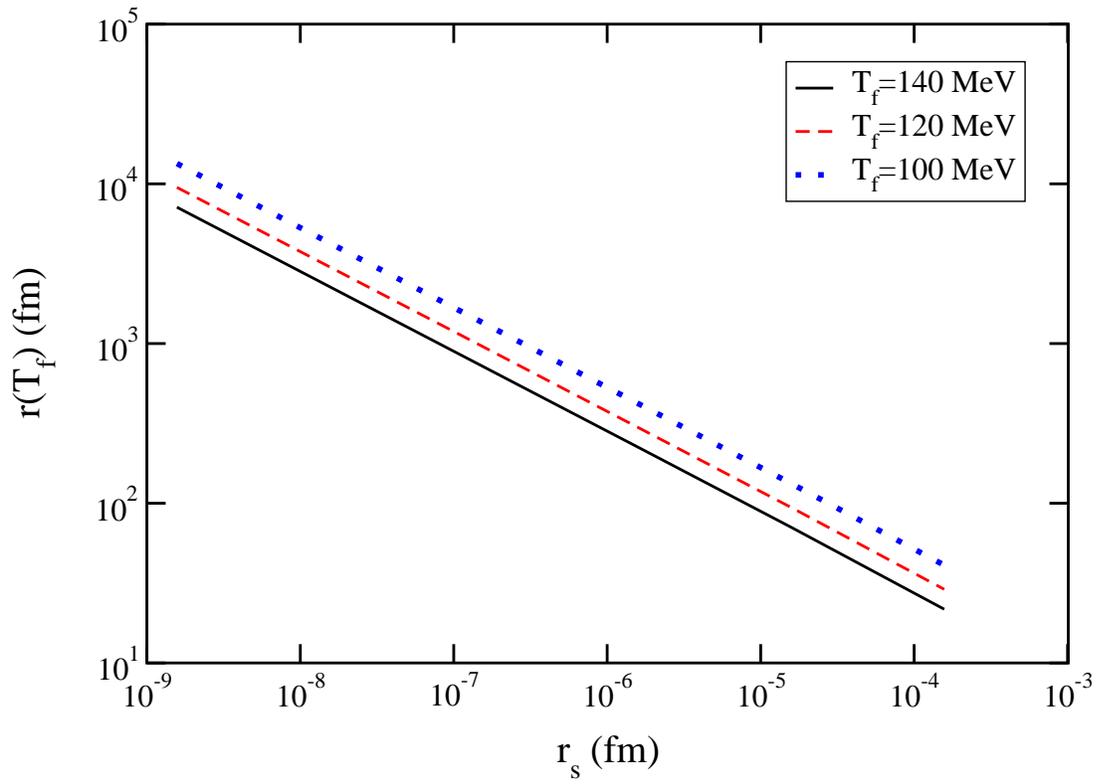,width=12.5cm,angle=270}}
\caption{The freeze-out or free-streaming radius as a function of the
Schwarzschild radius for three different freeze-out temperatures.}
\end{figure}

\begin{figure}
\centerline{\epsfig{figure=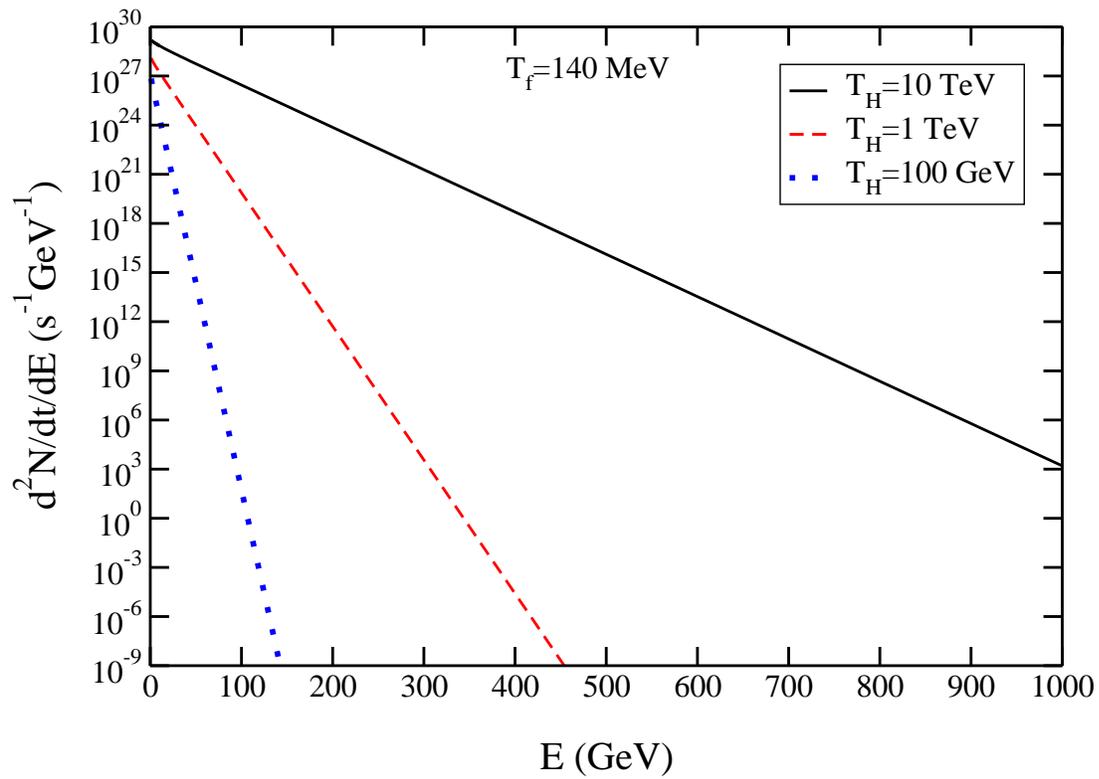,width=12.5cm,angle=270}}
\caption{The instantaneous gamma ray spectrum for three different
Hawking temperatures assuming $T_f = 140$ MeV.}
\end{figure}

\begin{figure}
\centerline{\epsfig{figure=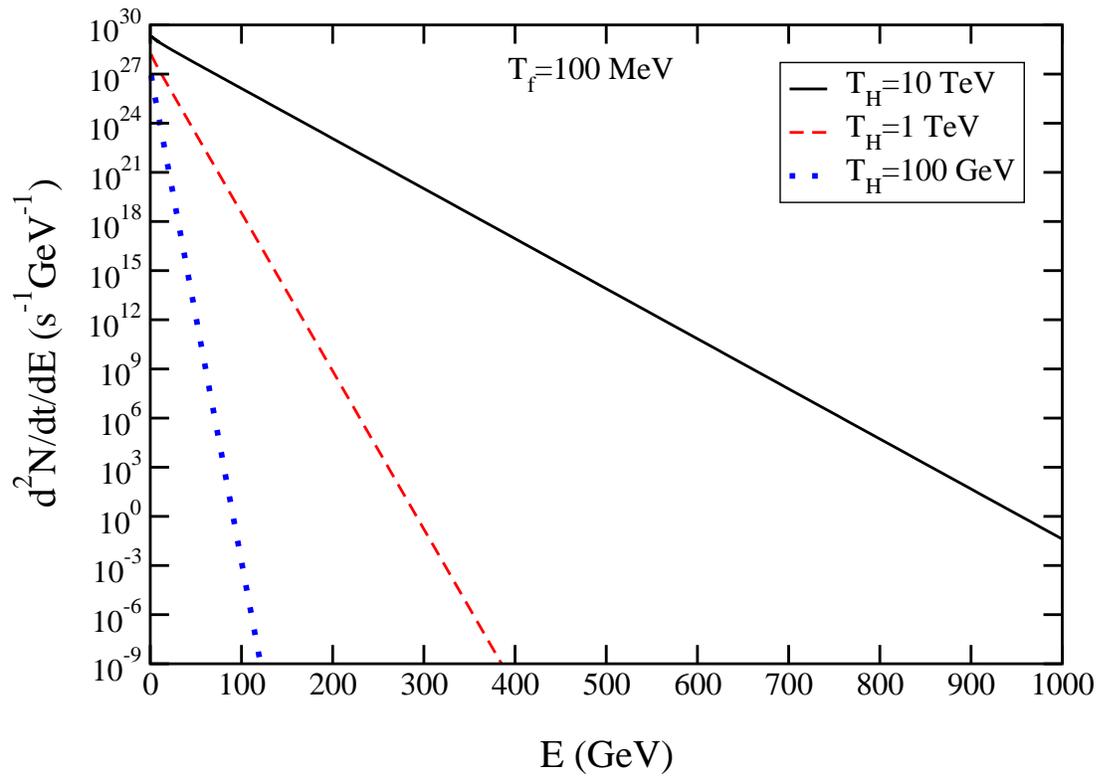,width=12.5cm,angle=270}}
\caption{Same as figure 9 but with $T_f = 100$ MeV.}
\end{figure}

\begin{figure}
\centerline{\epsfig{figure=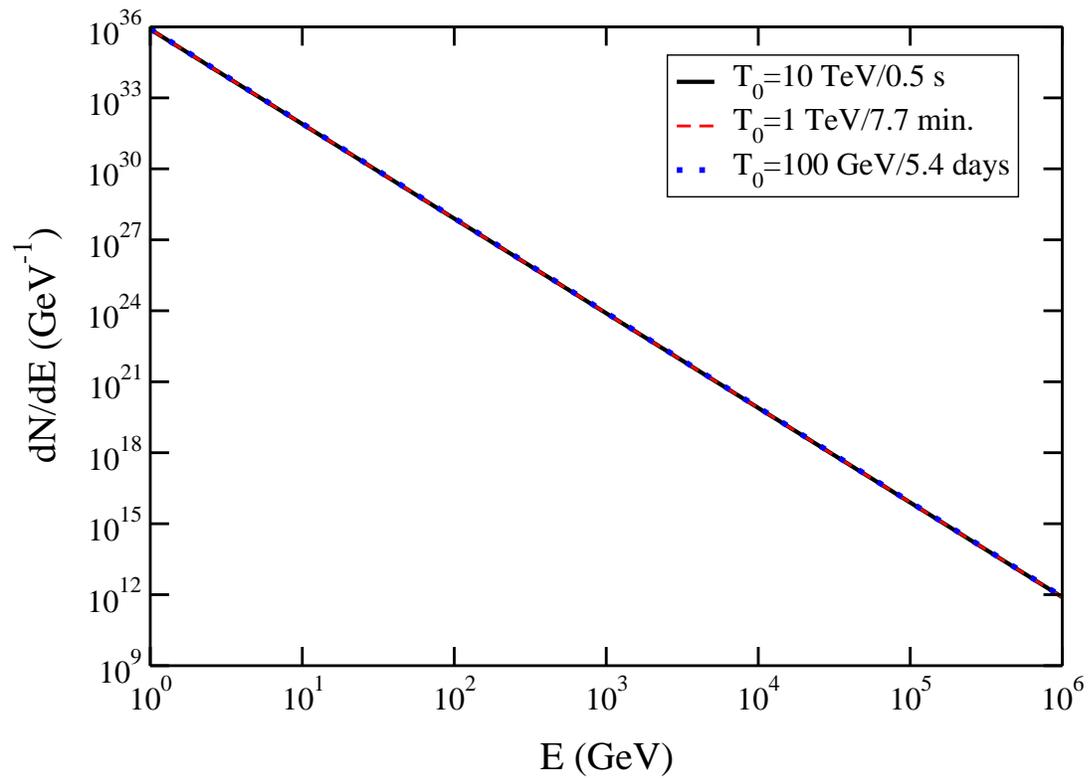,width=12.5cm,angle=270}}
\caption{The time-integrated gamma ray spectrum starting from the
indicated Hawking temperature.  Here $T_f = 140$ MeV.}
\end{figure}

\begin{figure}
\centerline{\epsfig{figure=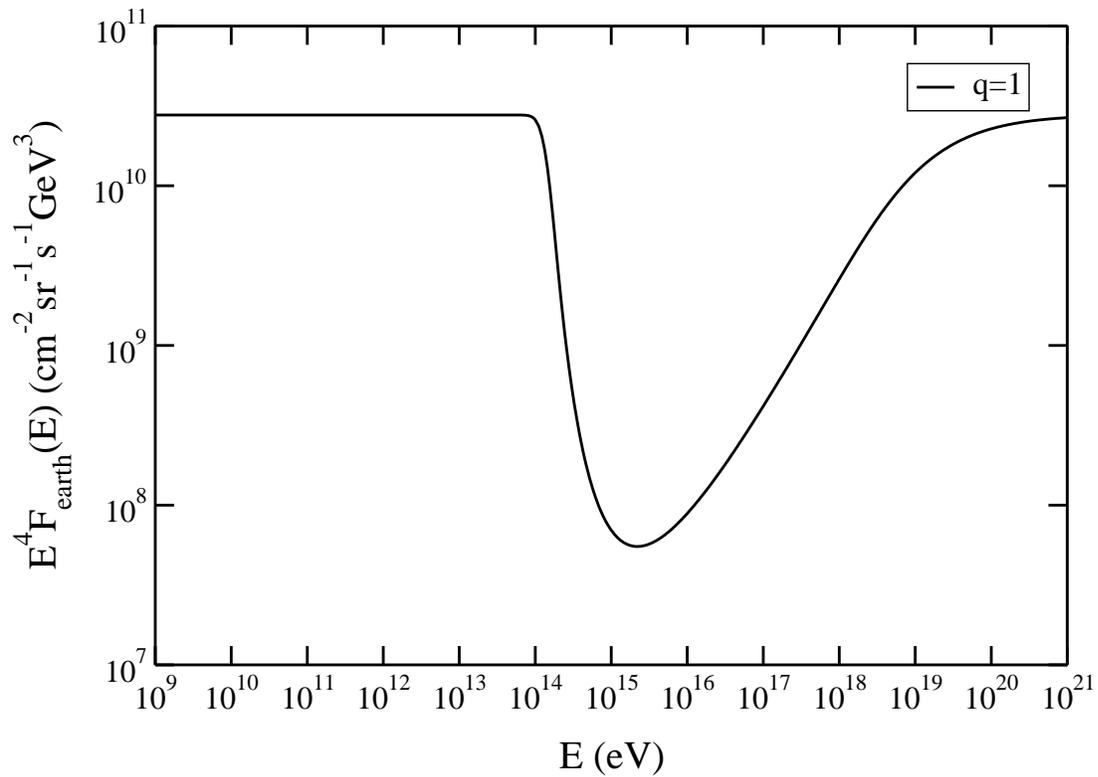,width=12.5cm,angle=270}}
\caption{The flux of diffuse gamma rays coming from our galactic halo.
The normalization is $\dot{\rho}_0 = 1$ pc$^{-3}$ yr$^{-1}$.
The halo is assumed to be spherically symmetric, $q = 1$; the results
for a flattened halo with $q = 2$ are very similar.}
\end{figure}

\begin{figure}
\centerline{\epsfig{figure=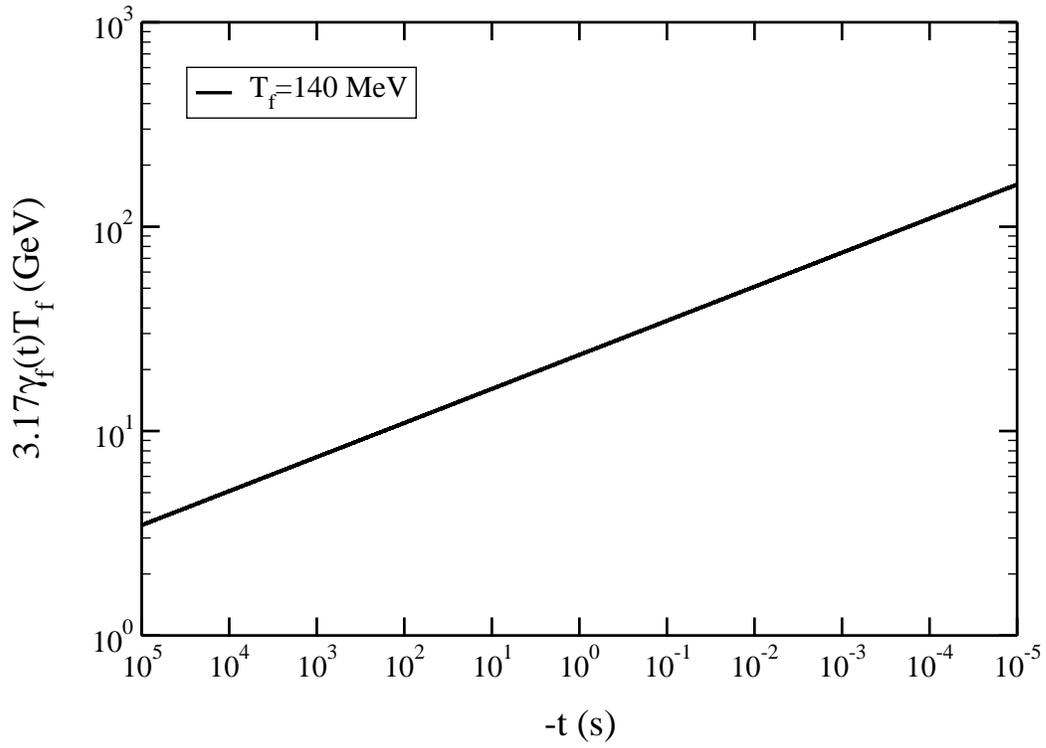,width=12.0cm,angle=270}}
\caption{The average gamma ray energy as a function of the remaining
lifetime of the black hole.  The times spanned correspond to approximately
400 GeV $< T_H <$ 200 TeV.}
\end{figure}

\end{document}